\documentclass[%
 reprint,
 amsmath,amssymb,
 aps,
]{revtex4-1}

\usepackage{graphicx}
\usepackage{dcolumn}
\usepackage{bm}

\makeatletter
\newcommand{\row}[1]%
{\mathord{\buildrel{\lower3pt%
\hbox{$\scriptscriptstyle\rightarrow$}}\over #1}}

\newcommand{\dyadic}[1]{\mathord{\dyadic@rrow{#1}}}
\newcommand{\dyadic@rrow}[1]{
\begin{picture}(12,12)(-1,0)
\put(-1,9){\makebox(0,0)[t]{$\scriptscriptstyle\downarrow$}}
\put(-1,9){\makebox(0,0)[l]{$\scriptscriptstyle\longrightarrow$}}
\put(5,0){\makebox(0,0)[b]{$#1$}}
\end{picture}
}

\newcommand{\bra}[1]{\bigl\langle #1 \bigr|}
\newcommand{\ket}[1]{\bigl| #1 \bigr\rangle}

\begin{document}

\preprint{APS/123-QED}

\title{Steering information in quantum network}

\author{N. Metwally}
 \altaffiliation{Department of Mathematics, College of Science, Bahrain
University, Bahrain.}
\footnotetext{This letter is submitted to third Smart city symposium , Bahrain 2020 }




\date{\today}

\begin{abstract}
In this contribution, we investigate the possibility that one member of  a quantum network can steer the information that encoded in the state of other member.   It is assumed that,  these  members   have a direct or indirect connections. We show that, the   steerability
 increases   at small values of the channel' strength.  Although, the degree of entanglement between the direct interacted nodes is smaller than that displayed for the non-interacted nodes, the possibilities  of steering a member of the direct interacted nodes and the non-direct nodes are almost similar.

\begin{description}
\item[Pacs] 03.67.Mn,~3.65.Yz,~03.67.Lx,~0.3.67.Bg
\end{description}

\end{abstract}

\pacs{Valid PACS appear here}
\maketitle


\section{\label{sec:Intro}Introduction}

\section{Introduction}
Handling information in the Smart cities is one of the most important characterize  the powerful  of the Smart cities' designing process.  To achieve this aim , one has to  generate  a secure network. New technologies could play central roles to keep the communication secure. In this context, we use the quantum network that generated by using Dzyaloshinskii-Moriya (DM) interaction \cite{Metwally2011,Metwally2018}.

Quantum steering is one of the fundamental concepts of quantum information \cite{Wiseman2007}. It has implemented for may different system theoretically and experimentally \cite{Jones2007}.  However, there are many studies have discuss this phenomena not only on inertial frame but  also in the non-inertial frame \cite{Weng}.  For example, the one way steering in the presences of thermal noisy is discussed by  Zhong et. al \cite{Zhong}.  Schneeloch et. al \cite{James2014} investigated the possibility of improving the EPR steering inequalities. In the critical systems, the EPR stering is studied by Cheng et. al \cite{Cheng2019}.

In this contribution, we investigate the  possibility that one member of this quantum network can steer another member, where we discuss whether   any nodes can detect the measurements done by his/her partner. We discuss this phenomena and its relation to the entanglement between each two connected nodes. Due to this type of interaction, there are two types of communication channels are generated, either by direct/indirect interaction. We investigate the effect of the integration strength on the degree of steerability.

\section{Quantum Network}
Quantum networks are consider as an alternative communication tool instead of the  well known. There are several protocols  are proposed for generating different types of quantum networks.
Let  us assume  that the suggested network consists of pairs of maximum entangled states of Bell states types;
$\ket{\phi^{\pm}}=\frac{1}{\sqrt{2}}(\ket{11}\pm\ket{00}),
\ket{\psi^{\pm}}=\frac{1}{\sqrt{2}}(\ket{10}\pm\ket{01})$ \cite{Metwally2014}.
 For example, the density operator
$\rho_{\phi^+}=\ket{\phi^+}\bra{\phi^+}$ takes the following form,

\begin{equation}\label{2Q}
\rho_{\phi^+}=\frac{1}{4}(1+\sigma_x\tau_x-\sigma_y\tau_y+\sigma_z\tau_z),
\end{equation}
where, the vectors $\row{\sigma_i}=(\sigma_x,\sigma_y,\sigma_z)$
and $\row{\tau_j}=(\tau_x,\tau_y,\tau_z)$ are Pauli operators (see
for example \cite{Metwally2011}).  Each particle represents a node on this quantum network, where they are  connected together
 via Dzyaloshinskii- Moriya (DM)
interaction \cite{You}, where the end of each entangled node interacts with
the first node of the other entangled two nodes. The describation of the suggested network is given in Fig.(1).  Let us assume that this quantum network consists four node. Therefore,
the initial state  of the network may be written as,
\begin{equation}\label{in}
\rho_{1234}(0)=\rho_{\phi^{+}_{12}}\otimes\rho_{\phi^{+}_{34}},
\end{equation}
where $\rho_{\phi_{12}^+}$ and $\rho_{\phi_{34}^+}$ are defined as
\begin{eqnarray}\label{2Q}
\rho_{\phi_{12}^+}&=&\frac{1}{4}(1+\sigma^{(1)}_x\tau^{(2)}_x-\sigma^{(1)}_y\tau^{(2)}_y+\sigma^{(1)}_z\tau^{(2)}_z),
\nonumber\\
\rho_{\phi_{34}^+}&=&\frac{1}{4}(1+\sigma^{(3)}_x\tau^{(4)}_x-\sigma^{(3)}_y\tau^{(4)}_y+\sigma^{(3)}_z\tau^{(4)}_z).
\end{eqnarray}
The  second and third nodes are connected via DM interaction,
which is defined by,
\begin{equation}\label{DM}
H_{DM}=\row{D}\cdot(\row{\sigma_i}\times\row{\tau_j}).
\end{equation}
The components of the vector  $\row{D}=(D_x,D_y,D_z)$ are the
strength of $DM$ interaction in the directions of $x,y$ and $z-$
axes  respectively \cite{Moh}.
If,  consider that DM is switched in $x-$ direction, then the time evolution of the initial network is given by \cite{Metwally2011}
\begin{equation}
\rho_{1234}^{final}=\mathcal{U}_x(t)\rho_{1234}(0)\mathcal{U}^{\dagger}_x(t)
\end{equation}
where $\mathcal{U}_x(t)=e^{-iH_{DM} t}$.

\begin{figure}[htp!]
	\centering
	\includegraphics[width=6.5cm, height=4.5cm]{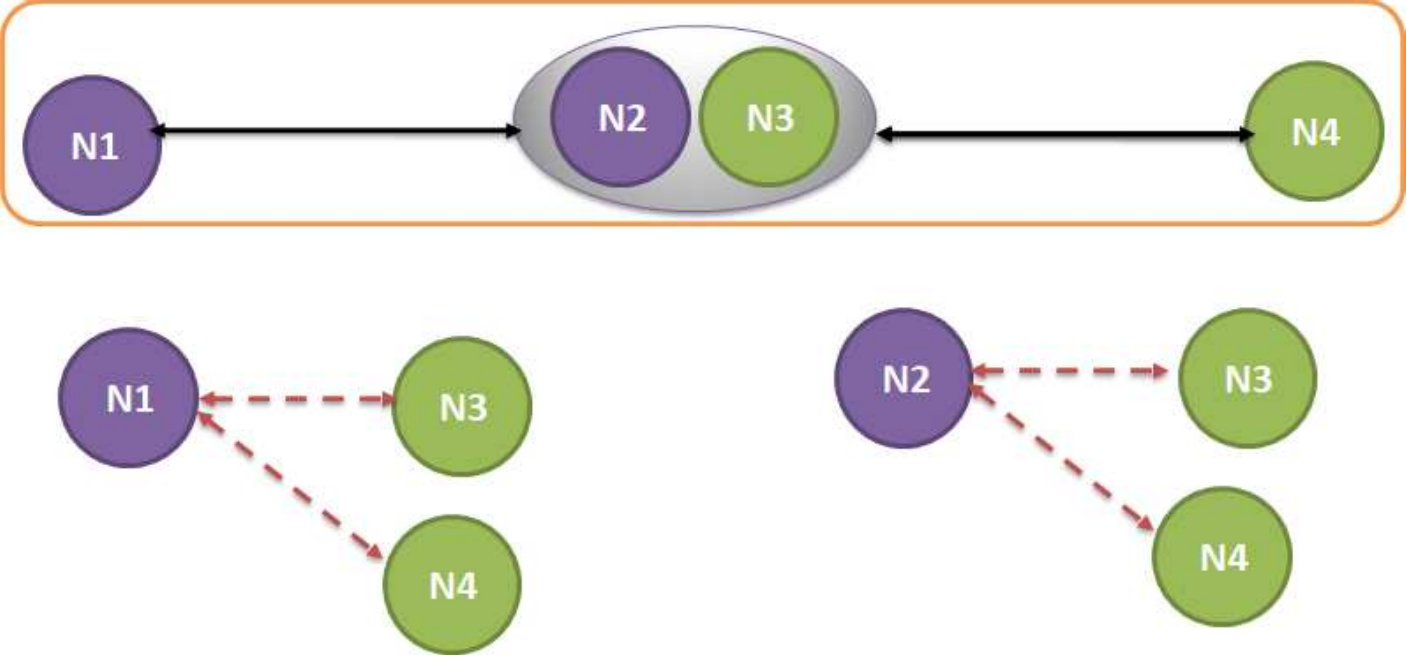}
\caption{The suggested entangled network, the nodes $N_1, N_2$  and $N_3, N_4$ share  maximum entangled states. The nodes $N_2$ and $N_3$ interact directly via Dzyaloshinskii- Moriya and consequently entangled state is generated between the second and the third nodes, $\rho_{23}$ directly, Moreover there are entangled  generated between the second and the fourth nodes, $\rho_{24}$, between the first and the third nodes, $\rho_{13}$ or between the first and the fourth nodes, $\rho_{14}$.}
 \end{figure}

Due to the interaction the four users  represent a quantum network consists of four nodes. The main task of this contribution is investigating the possibility of allowing one user to steer the state of another users.
From the final state $\rho_{1234}^{final}$, one gets the following partitions:$\rho_{23}=Tr_{14}(\rho_{1234}^{final})$, which represents direct interaction, while the indirect states are defined by $\rho_{14}=Tr_{23}(\rho_{1234}^{final})$ and $\rho_{13}=Tr_{24}(\rho_{1234}^{final})$. In an explicit form, these states may be written as,
\begin{eqnarray}
\rho_{23}&=&\mathcal{B}_1*I_{2\times 2}
\nonumber\\
&+&\mathcal{B}_2\Bigl(\ket{01}\bra{10}+\ket{10}\bra{01}\Bigr)
\nonumber\\
&+&\mathcal{B}_3\Bigl(\ket{11}\bra{00}+\ket{00}\bra{11}\Bigr).
\end{eqnarray}
where,
\begin{eqnarray}
\mathcal{B}_1&=&\frac{1}{4}\Bigl(\cos^4D_xt+\sin^4D_xt+\sin 2D_x t\Bigr)
\nonumber\\
\mathcal{B}_2&=&\frac{1}{2}\cos^{2}2D_xt, \quad \mathcal{B}_3=\frac{1}{8}\sin{^2}2D_xt.
\end{eqnarray}

The non-direct interacted nodes is defined by  either  $\rho_{24}$ or  $\rho_{13}$. In this context, we consider the state that is generated between the second and the fourth nodes which is given by,
\begin{eqnarray}
\rho_{24}&=&\mathcal{T}_1\times I_{4\times 4}
+\mathcal{T}_2\Bigl(\ket{00}\bra{11}+\ket{01}\bra{10}
\nonumber\\
&&+\ket{10}\bra{01}+\ket{11}\bra{00}\Bigr),
\end{eqnarray}
where
\begin{eqnarray}
\mathcal{T}_1&=&\frac{1}{4}(1+\sin^{2}2D_xt),~ \mathcal{T}_2=\frac{3}{8}(1+\sin^{2}2D_xt)
\nonumber\\
\end{eqnarray}

\subsection{Inequality of steerability and entanglement}
For a two qubit system  $\rho_{ab}$, the  possibility  that one  user $a$ steers the qubit $b$ if the following inequality is  satisfied,\cite{James2014,Wen2017}.
\begin{equation}
\sum_i\Bigl\{\mathcal{H}(\sigma^{(a)}_i|\sigma^{(b)}_i)\Bigr\}<2, ~\quad i=x,y,z
\end{equation}
where the users perform the Pauli $\sigma_x, \sigma_y$ and $\sigma_z$, measurements,  and $\mathcal{H}(A|B)=\mathcal{H}(\rho_{ab})-\mathcal{H}(\rho_a)$ is the conditional von-Neumann entropy. If we consider the state $\rho_{\ell m}$, then the post  measurement with respect to the  Pauli measurements   is  given by
\begin{eqnarray}
\tilde\rho_{\ell m}=\sum_{i=x,y,z}^{j}(\ket{\psi_i^{j}}\bra{\psi_i^{j}}\otimes I)\rho_{\ell m}(\ket{\psi_i^{j}}\bra{\psi_i^{j}}\otimes I)
\end{eqnarray}
where $j=1,2$ and $\ket{\psi_i^{j}}\bra{\psi_i^{j}}$ are the eigenvectors of the operators $\sigma_i, i=x,y,z$.
The degree of entanglement that is generated between the different nods is quantify by using the negativity\cite{Horodecki}. This measure based on the eigenvalues of the partial transpose of the state between the nodes. For any bipartite state $\rho_{ab}$, the negativity $\mathcal{N}(\rho_{ab})$ is defined as,
\begin{equation}
\mathcal{N}(\rho_{ab})=\sum_{i=1}^{4}\lambda_i-1,
\end{equation}
where $\lambda_i$ are the eigenvalues of $\rho_{ab}^{T_b}$ and $0\leq \mathcal{N}(\rho_{ab})\leq 1$.
\subsection{Direct generated state}
Due to the interaction, between  there is an entangled is generated  say $\rho_{23}$. Then the post measurement for this state with respect to Pauli-operator $\sigma_x$ is defined by , $\tilde\rho^{x}_{23}$ , where
\begin{eqnarray}
\tilde\rho^{(x)}_{23}&=&\mathcal{B}_1\times I_{4\times 4}
\nonumber\\
&&+(\mathcal{B}_{2}+\mathcal{B}_{3})\Bigl(\ket{00}\bra{11}+\ket{10}\bra{01}
\nonumber\\
&&+\ket{01}\bra{10}+\ket{11}\bra{00}\Bigr).
\end{eqnarray}
One can easily evaluate the  eigenvalues of the state $\tilde\rho^{(x)}_{23}$ and may be defined as $\lambda_n(\tilde\rho^{(x)}_{23}), n=1..4$.
Moreover, the  eigenvalues of the reduced density operator, $\tilde\rho_3^{(x)}=tr_2\{\tilde\rho^{x}_{23})\}$  may be described by $\mu_\ell(\tilde\rho_3^{(x)}), \ell=1,2.$
Similarly in the computational basis, the state $\tilde\rho^{(y)}_{23}$, may be written as
\begin{eqnarray}
\tilde\rho^{(y)}_{23}&=&\mathcal{B}_{1}\times I_{4\times 4}
\nonumber\\
&&+(\mathcal{B}_{3}-\mathcal{B}_{2})\Bigl(\ket{00}\bra{11}-\ket{10}\bra{01}
\nonumber\\
&&-\ket{01}\bra{10}+\ket{11}\bra{00}\Bigr).
\end{eqnarray}
one can simply evaluate the eigenvalues of this density operator which can be described $\lambda_i(\rho^{(y)}_{23}), i=1...4$.
Similarly the  eigenvalues of the reduced density operator $\tilde\rho_3^{(y)}=tr_2\{\tilde\rho^{y}_{23})\}$ are described by
$\mu_j(\tilde\rho_3^{(y)}), j=1,2$.
Finally,  the state $\tilde\rho^{z}_{23}$ takes the form
\begin{eqnarray}
\tilde\rho^{(z)}_{23}&=&\mathcal{B}_{1}\times I_{4\times 4}.
\end{eqnarray}
The reduced density of this state, $\tilde\rho_3^{(z)}=tr_2\{\tilde\rho^{z}_{23})\}$ has eigenvalues  $\mu_p(\tilde\rho_3^{(z)}), p=1,2$.

Then by using the eigenvalues of the three densities operator, $\tilde\rho^{(x)}_{23}, \tilde\rho^{(y)}_{23}$ and $\tilde\rho^{(z)}_{23}$ as well as the eigenvalues of their reduced density operators, one can easily evaluate the steering inequality (10).
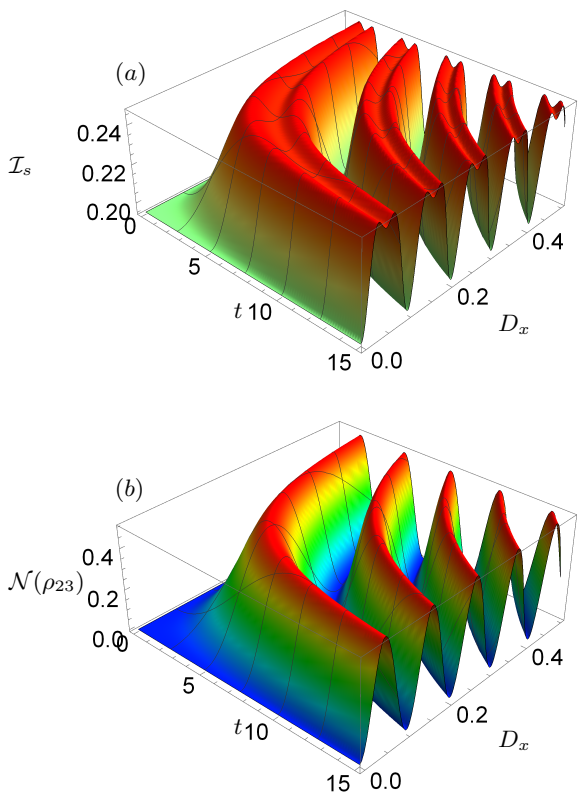
\begin{figure}[h!]
	\centering
	\includegraphics[width=0.8\linewidth, height=5.5cm]{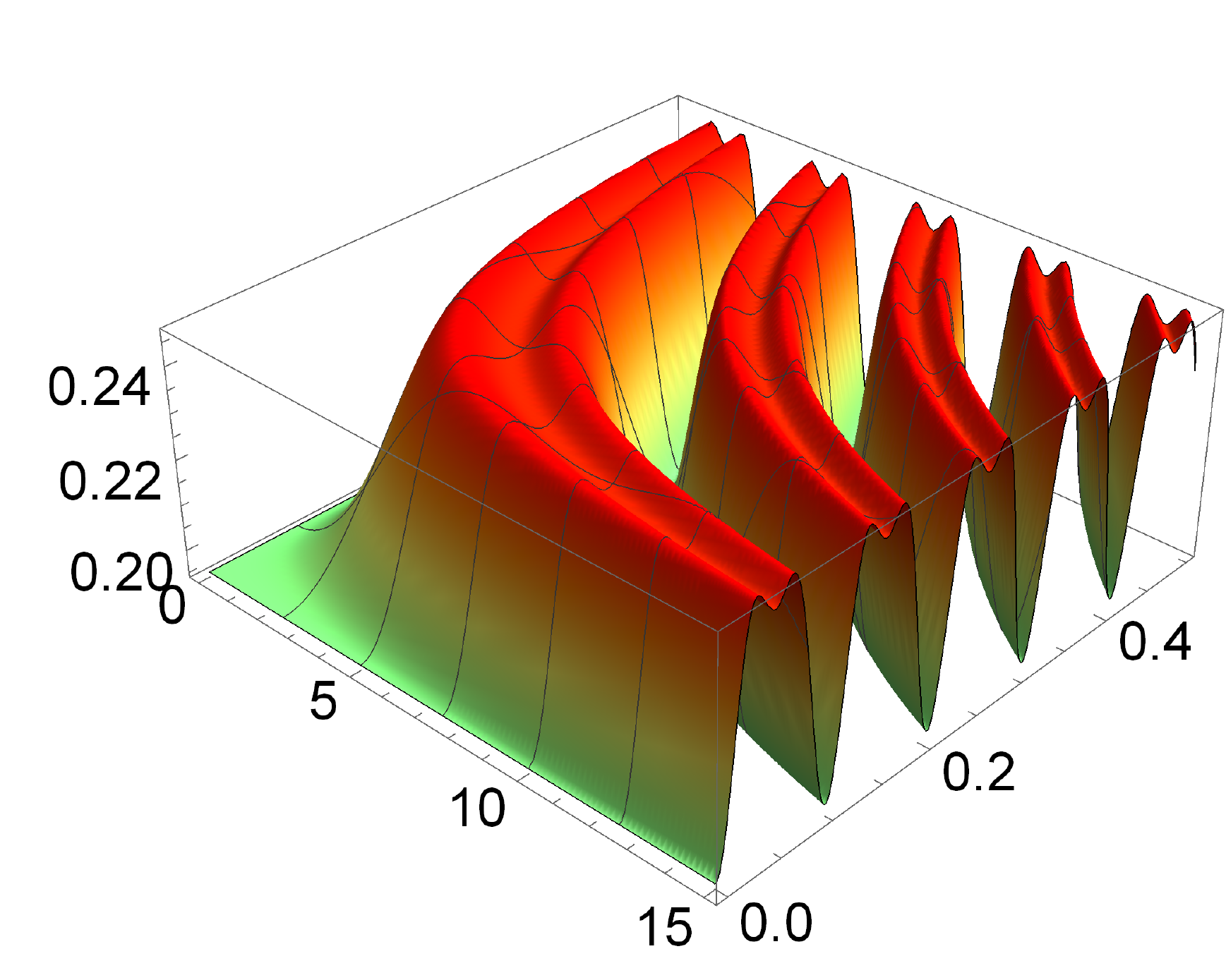}
 \put(-175,115){${(a)}$}
 \put(-30,20){${D_x}$}
  \put(-130,25){${t}$}
  \put(-215, 80){$\mathcal{I}_s$}\\
\includegraphics[width=0.8\linewidth, height=5.5cm]{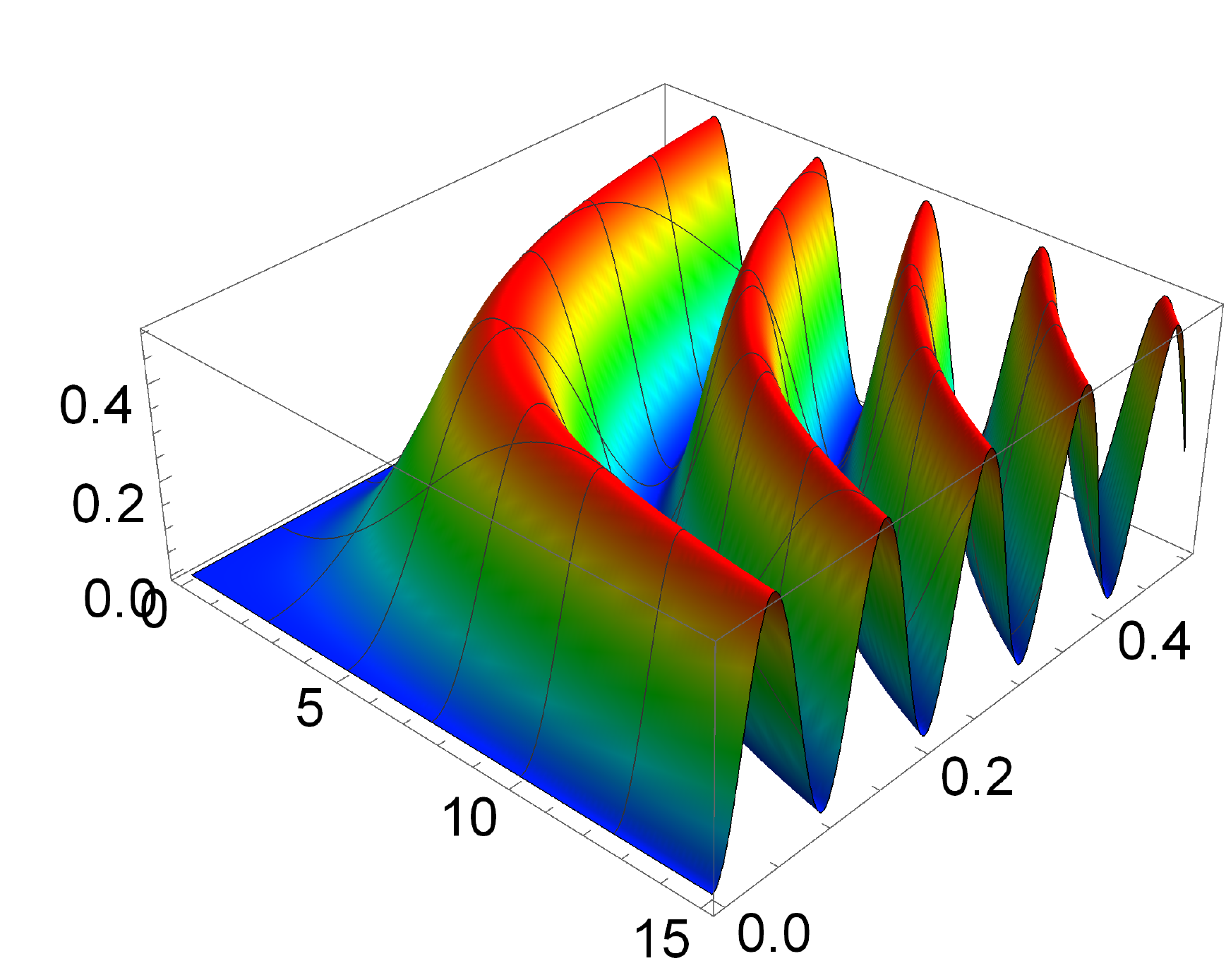}
 \put(-175,115){${(b)}$}
 \put(-30,20){${D_x}$}
 \put(-130,25){${t}$}
 \put(-215, 80){$\mathcal{N}(\rho_{23})$}
	\caption{(a)The behaviour of the quantum steering inequality(b) The degree of entanglement$\mathcal{N}(\rho_{23})$}
\end{figure}
The behavior of the steering inequality (10) is displayed in Fig.(2a). It is clear that, the  possibility that the Alice can steer the Bob'states at any given strength of the DM interaction changes periodically between maximum and minimum bounds, However, the steering inequality is obeyed,where $\mathcal{I}_{s}<2$. These means that any measurements performed by Alice, Bob can predicted it and consequently changes his state accordingly. These results may be confirmed from Fig.(2b), where the   amount of entanglement that generated between the two particles is quantified by using the negativity. The behavior of $\mathcal{N}(\rho_{23})$ shows that, the entanglement is generated between the two qubits as soon as the interaction is switched on. Moreover, the entanglement fluctuates between the upper and lower bounds.

\begin{figure}[h!]
	\centering
	\includegraphics[width=0.65\linewidth, height=4cm]{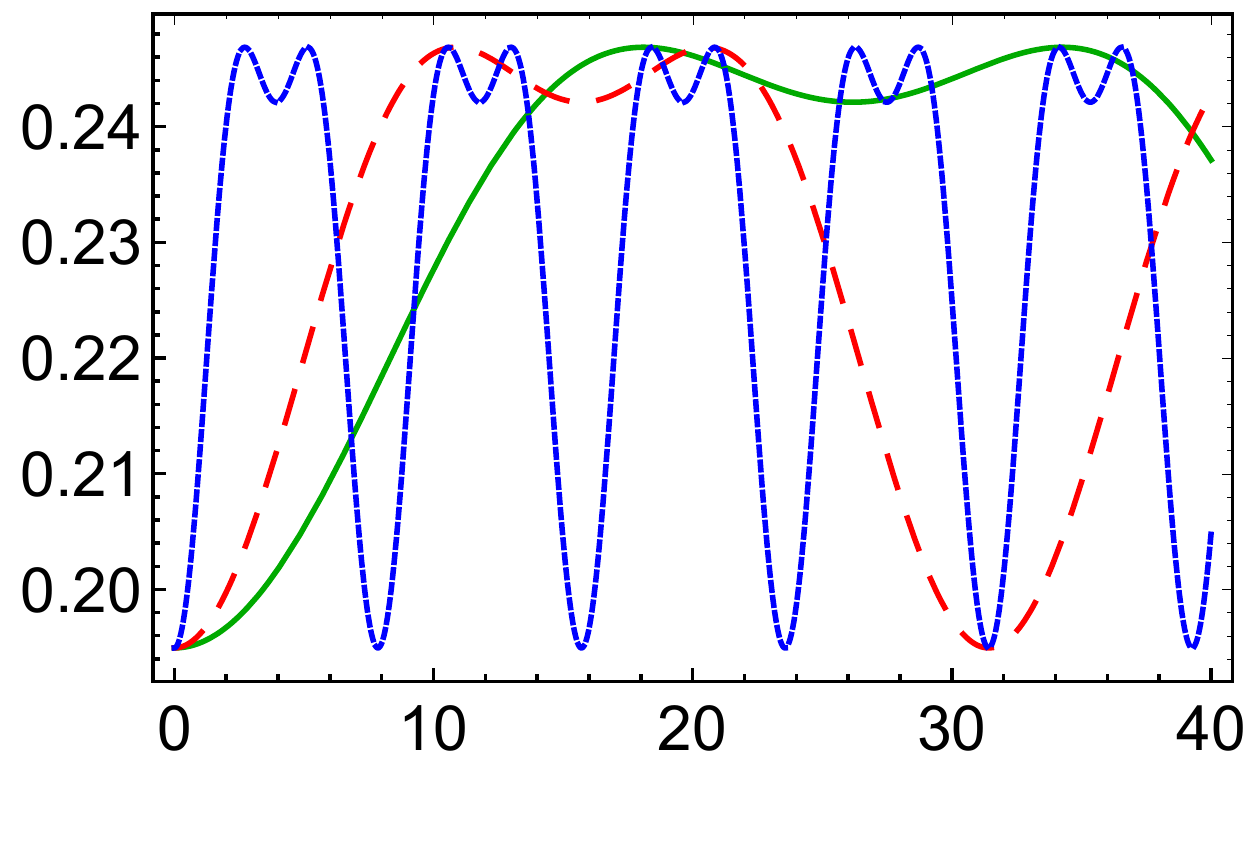}
 \put(-160,110){${(a)}$}
 \put(-75,5){${t}$}
    \put(-190, 70){$\mathcal{I}_s$}\\
\includegraphics[width=0.65\linewidth, height=4cm]{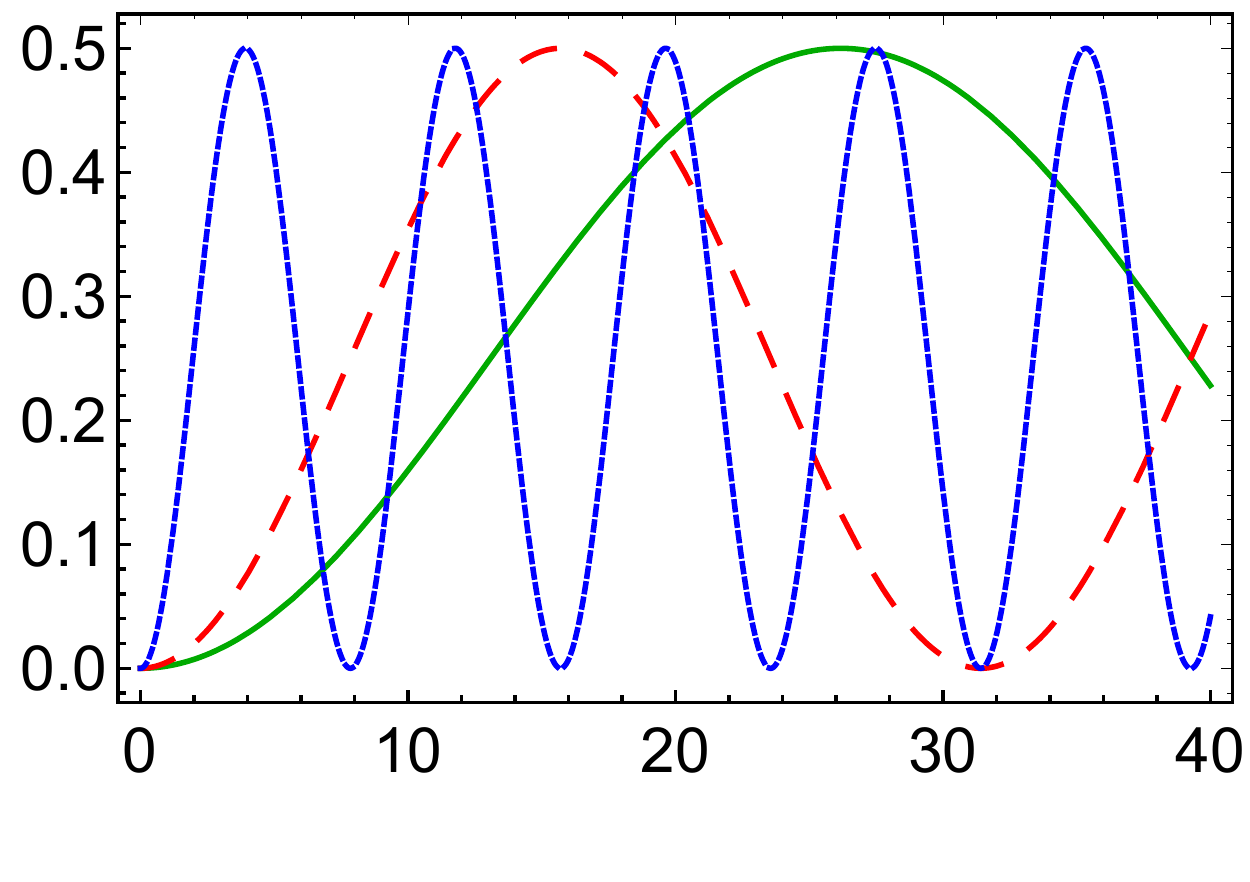}
 \put(-160,110){${(b)}$}
 \put(-75,5){${t}$}
 \put(-190, 70){$\mathcal{N}(\rho_{23})$}
	\caption{(a)The  quantum steering inequality, $\mathcal{I}_s$ (b) The degree of entanglement $\mathcal{N}(\rho_{23})$, where the  solid, dot and dash at $D_x=0.03,0.05$ and $0.1$, respectively.}
\end{figure}
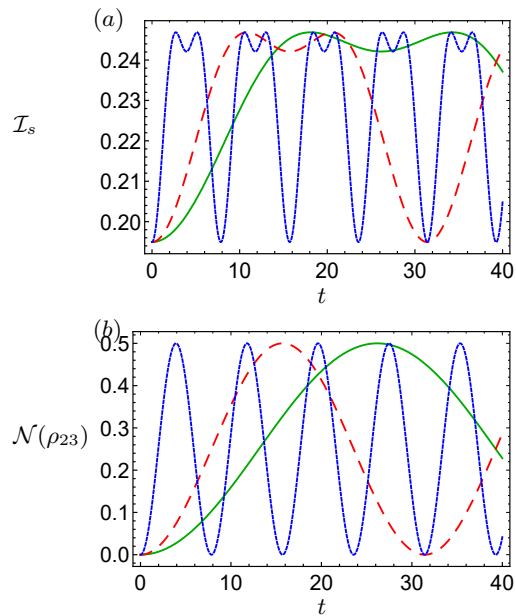

In Fig.(3) we discuss the behavior of the steering inequality and the entanglement at some different values of the interaction strength $D_z$.
It is clear that, the  system satisfies the  steering inequality, where $\mathcal{I}_s$ oscillates between  its maximum and minimum bounds. The number of oscillations increases as the interaction strength $D_z$ increases.  Moreover, the steering inequality increases  suddenly as the  interaction $D_Z$, while gradually behavior is predicted at smaller values of $D_z$.

\subsection{Indirect generated state}
In this subsestion, we investigate the steerability between the second and the fourth nodes, who share the state $\rho_{24}$ (Eq.(8)), which  represent the indirect entangled nodes. Then by using the   post selection measurements in the $x$-direction, one gets,
\begin{eqnarray}\label{rho24}
\tilde\rho^{(x)}_{24}&=&\rho_{24},
\nonumber\\
\tilde\rho^{(y)}_{24}&=&\tilde\rho^{(z)}_{24}=\mathcal{T}_1\otimes I_{4\times 4}.
\end{eqnarray}
Now, we have all the details to investigate the steerability phenomena, by using  the eigenvalues of the Eq.(\ref{rho24}) and their reduced density operators  in steering inequality (10).

\begin{figure}[h!]
	\centering
	\includegraphics[width=0.8\linewidth, height=5.5cm]{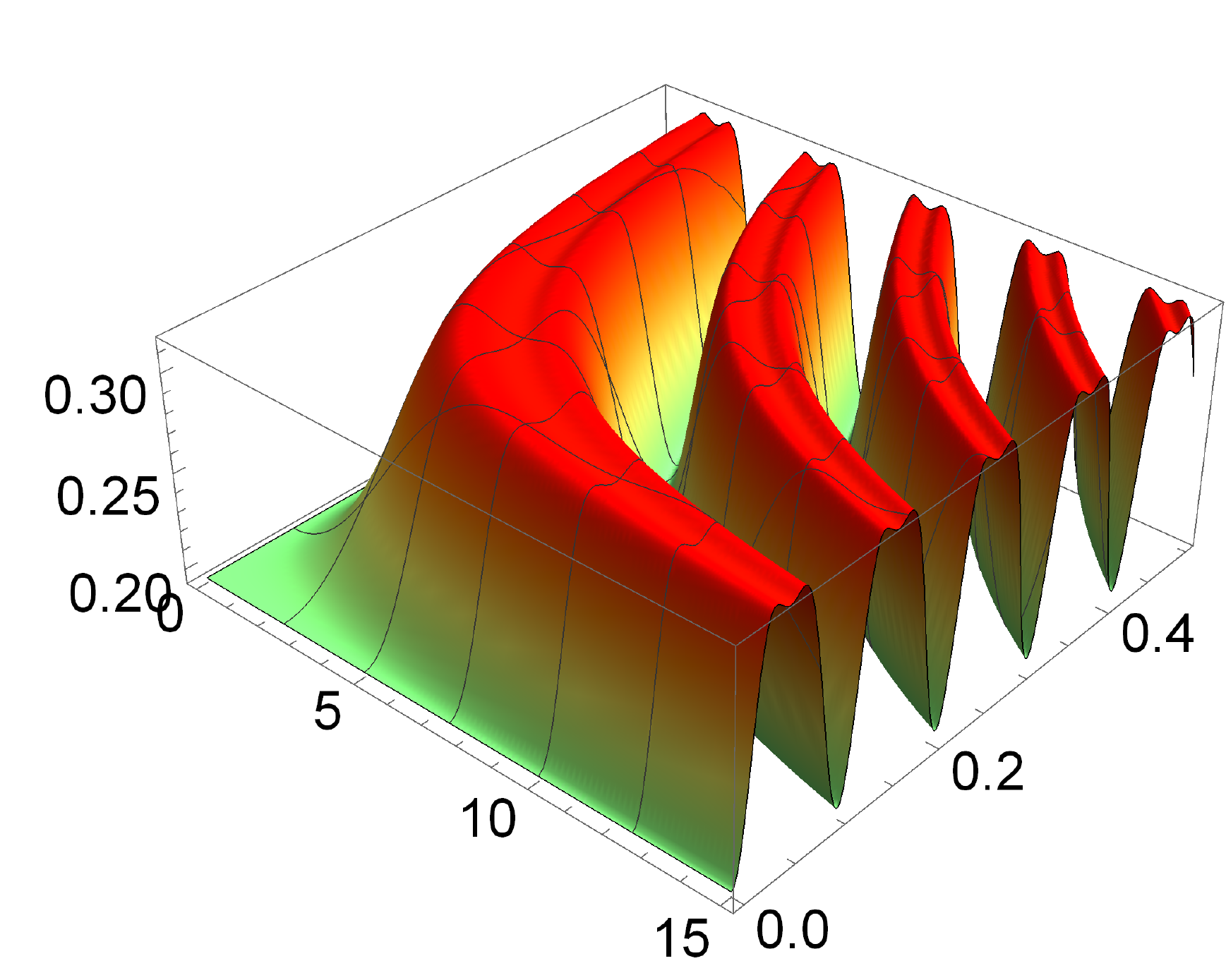}
 \put(-175,110){${(a)}$}
 \put(-30,20){${D_x}$}
  \put(-130,25){${t}$}
  \put(-210, 80){$\mathcal{I}_s$}\\
\includegraphics[width=0.8\linewidth, height=5.5cm]{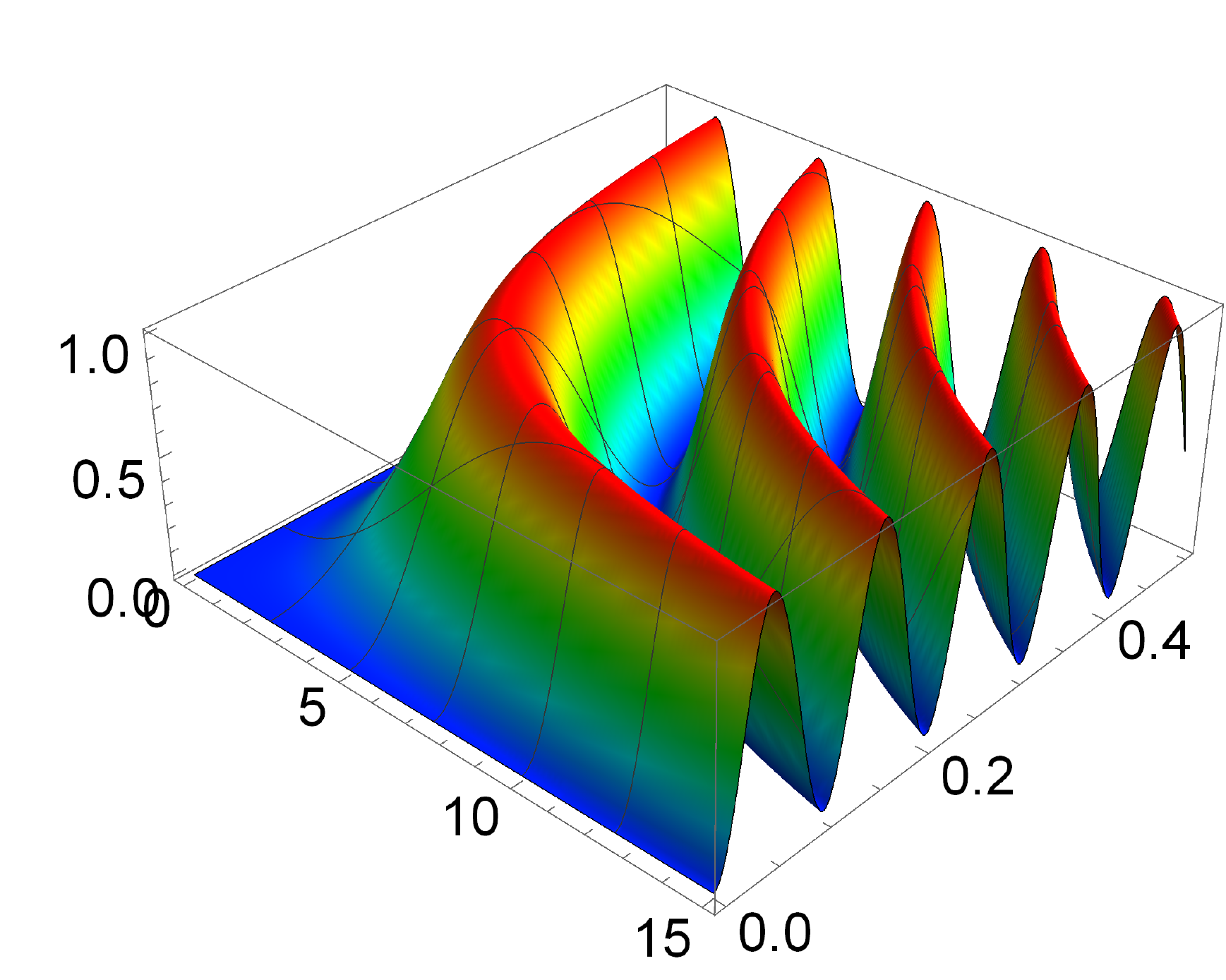}
 \put(-175,110){${(b)}$}
 \put(-30,20){${D_x}$}
 \put(-130,25){${t}$}
 \put(-215, 80){$\mathcal{N}(\rho_{24})$}
	\caption{The same as Fig.2 but for  a non direct interacted nodes, $\rho_{24}$.}
\end{figure}
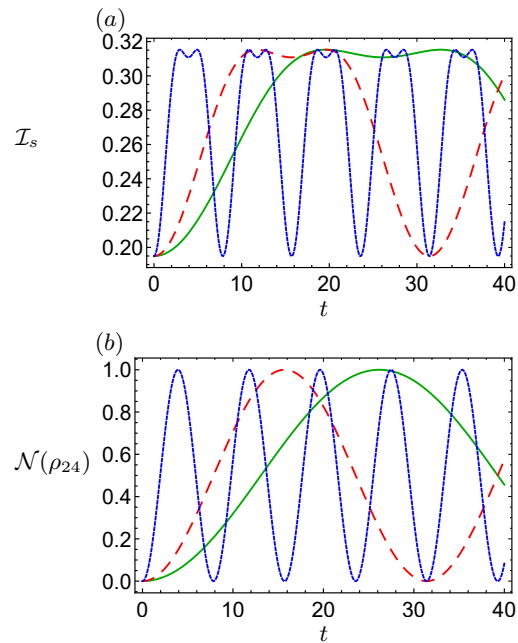
\begin{figure}[h!]
	\centering
	\includegraphics[width=0.65\linewidth, height=4cm]{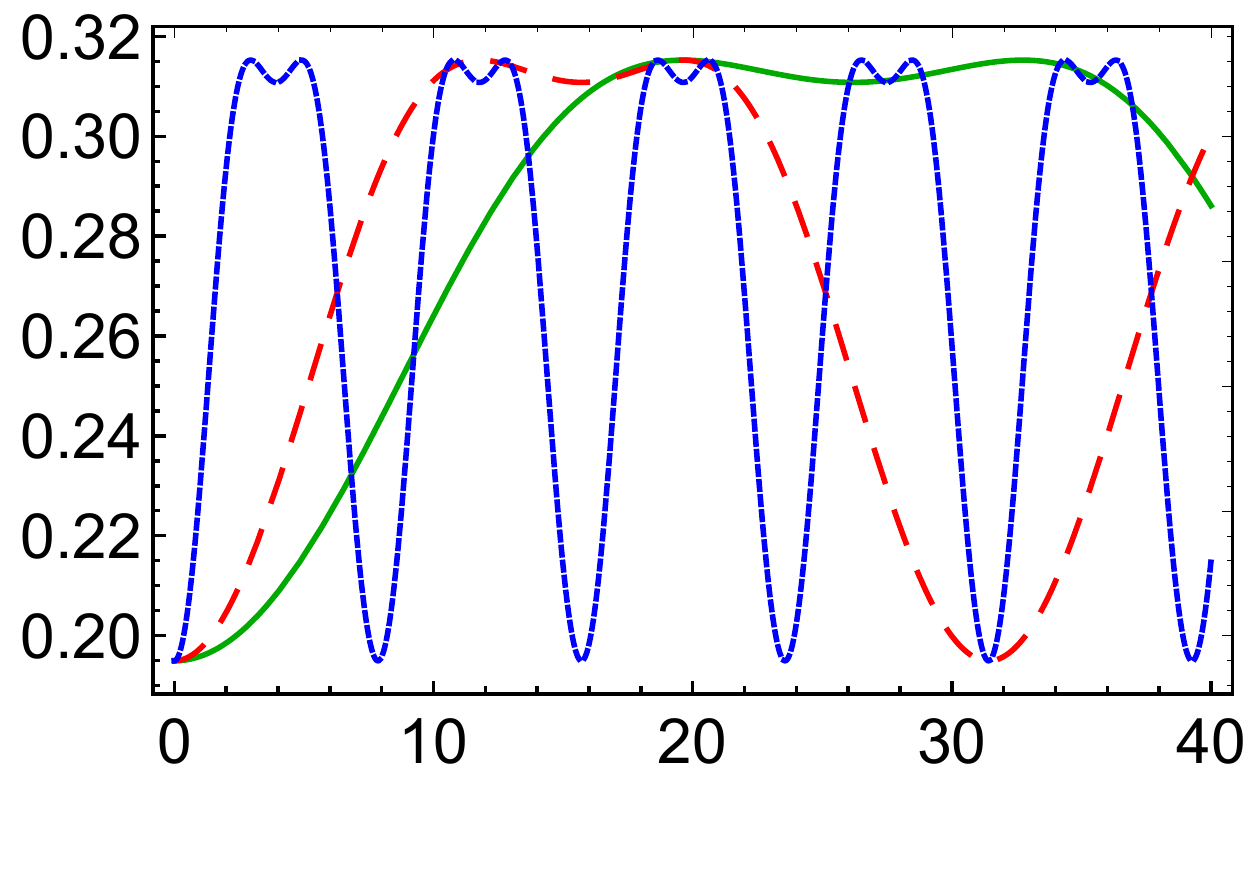}
 \put(-160,115){${(a)}$}
 \put(-75,5){${t}$}
    \put(-190, 70){$\mathcal{I}_s$}\\
\includegraphics[width=0.65\linewidth, height=4cm]{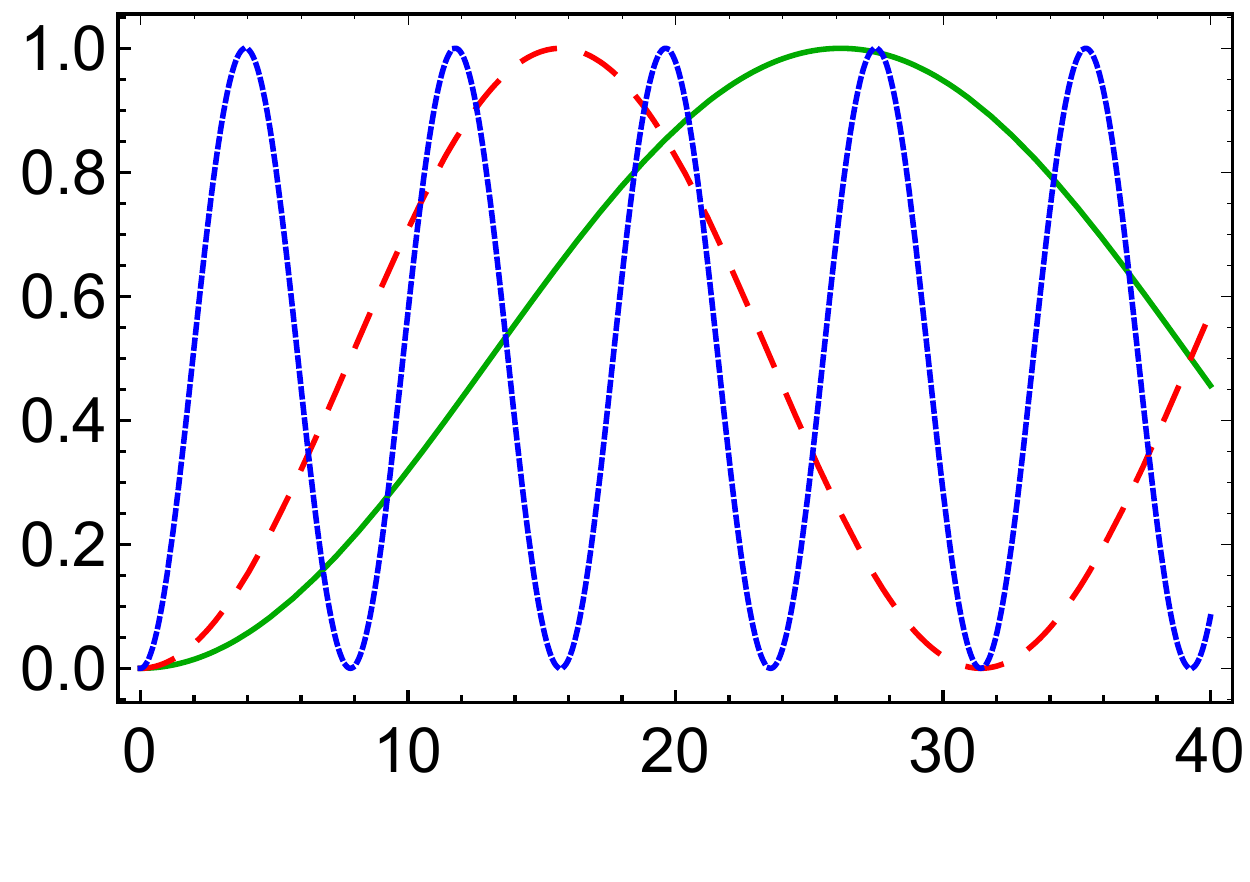}
 \put(-160,115){${(b)}$}
 \put(-75,5){${t}$}
 \put(-190, 70){$\mathcal{N}(\rho_{24})$}
	\caption{(a)The  quantum steering inequality, $\mathcal{I}_s$ (b) The degree of entanglement $\mathcal{N}(\rho_{24})$, where the  solid, dot and dash at $D_x=0.03,0.05$ and $0.1$, respectively.}
\end{figure}

In Fig.(5), we investigate the steerability for a non-directed interacting nodes where we consider the quantum channel between the second and the fourth nodes, $\rho_{24}$. The behavior is similar to that depicted for direct generated entangled channel. However as the entanglement vanishes, i.e., the two nodes are separable, the steerability is zero. It is clear that, the maximum bounds of the entanglement are larger than those displayed in Fig.(3b).  Moreover, the small values of the DM interaction increase the  possibility steerability.

In Fig.(5b), we discuss the steerability at some values of the integration strength, the behavior is similar to that displayed in Fig.(4a). The  interval  of time in which the steerability is predicted increases  at small values of the interaction' strength. Moreover, the possibility that the steerable node can predict the steerer's measurements is almost fixed at large interval of time.

\section{Conclusions}
In this contribution, we investigate the possibility of steering one qubit in a quantum network, where this network is generated by using the Dzyaloshinskii-Moriya (DM) interaction. It is clear that, the  steerability  of one particle in quantum network is possible between the direct connect nodes at any value of the  interaction strength.  These results, coincide with the  behavior of entanglement, where as soon as the integration is switched on, an entangled state is generated and consequently one member of the quantum network can steer another member.  The steerability  vanishes as soon as the entangled nodes turn into separable nodes.   The steerability increases as  at small values of the channel' strength, where the  steerability periodic of time increases.  Although, the degree of entanglement between the direct interacted nodes is smaller than that displayed for the non-interacted nodes, the possibilities  of steering a member of the direct interacted nodes and the non-direct nodes are almost similar.


\end{document}